\documentclass{bull-l}

\theoremstyle{definition}

\theoremstyle{remark}

\numberwithin{equation}{section}

\begin{document}

\title{Proof in the time of machines}
\author{Andrew Granville}

 \address{D{\'e}partment  de Math{\'e}matiques et Statistique,   Universit{\'e} de Montr{\'e}al, CP 6128 succ Centre-Ville, Montr{\'e}al, QC  H3C 3J7, Canada.}
   \email{andrew.granville@umontreal.ca}  
\subjclass[2020]{Primary }

\date{}
\begin{abstract}
We compare the values associated with (traditional) community based proof verification to those associated with computer  proof verification. We propose ways that computer proofs might incorporate successful strategies from human experiences.
\end{abstract}

\maketitle 

\section{Introduction}

Professional pure mathematicians are complicit in the undergraduate fantasy that mathematics is solidly built up from a bedrock of axioms, into an immutable structure of objectively proven true results.\footnote{A beautiful conceit that persuades generations of undergraduates that mathematics is beautiful and timeless.}  However any experienced
working researcher proceeds rather differently, building on a  library of knowledge (as found in papers, preprints and books, as well as tacit knowledge) which they assume and hope is correct (or at least correctable).
In practice \emph{formal proofs}, chasing a proof back to the axioms, seem both tiresome and more-or-less pointless since progress is mostly measured in what is new and added to our knowledge, with reference to the distant horizon, not by a tedious and hard-to-follow level of rigour, safety checks on what is far behind us. On the other hand \emph{intuitive proofs} not only provide acceptable justification of claims to a community of researchers (at least for now), but also stimulate, motivate and inspire the next generation of developments.

However, change is afoot.  With the advent of usable proof verification systems, with a large library of already given definitions, together with a wide panorama of fundamental theorems inexorably deduced from axioms, a researcher might hope to have a formal proof verification available by inputting their standard \emph{intuitive} proof (albeit with modifications to fit what the proof system knows and needs) and feel more confident that they have not missed some simple issue.

Moreover once machines can standardly convert an intuitive proof into a formal proof, and vice-versa, surely machines can help in the process of proof, at first confirming proposed lemmas (by recognizing  proofs that are already in their library, or finding straightforward modifications of what is already there). At some point we might expect such machine proof-verifiers to be used as `black boxes'', in which we won't need to understand what they're doing to believe they have found an objective proof. Much of pure mathematics research  is focused on proofs of proposed statements, and it is hard to guess whether there will be any limits to machines' participation in this process?
This raises many questions about the future of research mathematics, and the role of humans in that future, as highlighted by Venkatesh's essay \cite{Venk}. Here I will focus on one important aspect: Will we have more confidence  that a machine-assisted formal proof is   objectively proven than the current more intuitive proofs. And if so, on what basis?
To properly consider such questions we should first perhaps appreciate what the current system of verifying proofs is and what advantages it has (for more details on many of these points see \cite{Gr}).
We should also better understand what  machine proof-verifiers actually do, including the role of the human team inputing the proof into the machine.

 \subsection*{A brief history of proof} Mathematicians \emph{prove} new statements by reducing them to those that have already been accepted as   true:
 \begin{quote}
If ... understanding is as we posited, it is necessary for demonstrative understanding ... to depend on things which are true and primitive and immediate and more familiar than and prior to and explanatory of the conclusion.  \mbox{} \hfill ---  \textsc{ Aristotle}
\end{quote}
This recursive process  must have  some starting point(s), some set of \emph{axioms}, ideally elegantly formulated, as few as possible and consistent (we shouldn't be able to justify a statement and its negation from our axioms!) to start the proving.  Moreover, it would be best if the axioms  allow us to prove all the theorems we are interested  in, and if not we should always be able to formulate a new axiom to help.

As far as we know, Euclid made the first serious attempt to formulate axioms and these were refined and developed well into the 20th century. Cantor in particular argued that mathematical progress depends on conceptual innovation so we should always be ready  to adapt our axioms, which should be \emph{evidently true} and consistent.  However Frege's widely touted system was inconsistent (Russell created paradoxes within the interpretation of its language), and then Russell and Whitehead's replacement was not entirely ``self-evident'' and unwieldy to implement.

These issues led Hilbert to suggest that we should be able to \emph{start from any set of consistent axioms} (with a well-defined simple language, and appropriate inference rules for proofs) and see where that leads; in other words, one does not have to start from immortal truths. It is the proofs that conjure the mathematics into existence. Frege disagreed, arguing that a line is a line, the physical entity in the common vernacular, and if you produce a theory yielding something different with an obscure though consistent set of axioms, then you should discard that  theory.

Hilbert's rules inevitably lead to a plurality of axiomatic systems: for example,   whether   there is an infinity in-between the infinity given by the integers and the infinity given by the reals, in size, is independent of the 10 ``standard axioms'' (ZFC), as proved by G\"odel and Cohen. 
Allowing such alternative foundations seems to reduce mathematics to a game with certain rules, rather than developments based on profound insights, but this is where mathematicians today apply Cantor's rules, that we only accept possibilities that lead to interesting innovation.

Hilbert's hope was to find a set of consistent axioms that are complete (that allow us to prove all theorems), and provably so. However in 1931 Kurt G\"odel shook Hilbert's proposed mathematics  to its foundations with his \emph{incompleteness theorems} which essentially say (simplifying crudely) that no consistent finite set of axioms and rules
can be used to prove all theorems about the integers, nor can they prove themselves to be consistent. Disaster. G\"odel's theorems directly imply that
\begin{quote} 
There can be no rigorous justification for classical mathematics.   \\ \mbox{} \hfill ---  \textsc{  John von Neumann}
\end{quote} 

\subsection*{Crisis response}
Most pure mathematicians deal with this irresolvable foundational crisis by ignoring it.\footnote{We occupy the penthouse of a very high, seemingly solidly built, tower, constructed over many generations. The engineers have informed us that   a  fire in the basement   has caused irreparable structural damage, and recommend demolition.  However they have no good plans for rebuilding to anything like the heights  already reached.  So instead we are  hoping that, now the smoke and noise have mostly dissipated and we have got used to the incessant alarm,  we can  continue on as if nothing is amiss, adding new floors and not worrying about what is going on below.} 
For many mathematicians, G\"odel's objections seem to be irrelevant to what they are working on. They need a
formal reasoning  system that is reliable in all ``reasonable circumstances'' (as in  the recent re-births of category theory). For the last century the basic axiomatic system  (ZFC) has remained accepted as the essentially unchanged foundation for most of modern pure mathematics,   despite G\"odel's results, and yet it works. But there is no end-run around G\"odel's theorems: various authors mistakenly believe that computers will somehow help, but that is without understanding the \emph{Church-Turing thesis}. This is the working hypothesis that all sensible computational systems are ``equivalent'' and ``universal''. This implies that they can each calculate anything that is calculable. Therefore they can imitate one another, and computers can perfectly imitate humans and human interaction.\footnote{In the imitation game \cite{Tur1} an interrogator tries to distinguish between  a human who tries to prove she is a human and    a computer who tries to fool the interrogator into believing it is a woman. (In the original, pre-Turing, game the part of the computer is played by a mischievous male.) A computer that can imitate humans so perfectly  will have passed the ``Turing test''.} Moreover humans can perfectly imitate computers, so  that our fundamental limitations are the same.
\begin{quote} 
 Digital computers \dots are intended to carry out any operations which could be done by a human. \hfill ---  \textsc{ Alan Turing}
\end{quote}

There is a less well-appreciated foundational crisis. Simple counting arguments establish that NP-families of provable correct statements (those for which there is a proof of ``polynomial size'' in the length of the problem)  form a rather tiny subset of the set of all families of provable correct statements in ZFC. Thus in practice, no machine, no algorithm, no human or computer can hope to prove all, or even a tiny fraction of, the  correct and provable mathematical statements deducible from this or any other plausible axiomatic system.\footnote{There is a secondary issue -- is there an algorithm that in polynomial time can find proofs for the theorems in a given NP-family?  If so, 
\begin{quote} If P$=$NP  then everything we are trying to do can be done ... However, I think most people believe that  P$\ne$NP.  \hfill ---  \textsc{ Avi Wigderson}.
\end{quote} }

\section{Accepting a result, by being convinced by a proof}

We would like to believe that pure mathematics has objective standards of proof, that any theorem has been rigorously justified back to the axioms.
But who verifies this, and how? One would like a highly-skilled objective verifier, a trusted authority, who can verify that the language and deduction rules have been validly used all the way back to the axioms. In a formal proof every inference is laid out so that the proof can be mechanically verified, requiring no intuition.\footnote{Though this ideal pre-supposes consensus on the meaning and interpretation of each step, and the inviolability of logic and language on the part of the verifier.}
But would such a proof be convincing to mathematicians? 
\begin{quote} 
Proof \dots supposedly establishes the undeniable truth of a piece of mathematics, [but] proof doesn't actually convince mathematicians of that truth ... Something else does.
 ---  \textsc{ Eugenia Cheng   \cite{Chen} }
\end{quote}
She claims that  mathematical communication  turns the author's beliefs  into a \emph{believed truth} of her reader via \emph{plausible reasoning}. This reasoning takes the structure of a carefully worded intuitive proof, with enough deductive  rigour to allay any fear of ambiguity or misdirection. It therefore fits into the mathematical community's perception of what is known, and what should be, suggesting a  verifier should fit any new claims into a larger context to help the reader.

Indeed, for highly sophisticated advances, the mathematical community accepts proofs if experts can fully appreciate the overall strategy, and then if each part of the proof can be verified by an expert on that technique.
Moreover the process of verification inseminates ideas in to   the research community: Verifiers are rarely looking just to agree, rather they are looking to incorporate new ideas and techniques into their own repertoire.
They do not do this by simply reading the words of the author, but rather by re-interpreting the text (or lecture) into their own terms, and matching up the new ideas with what they  know or believe.

\subsection*{Believing a proof, a community perspective} We believe  a proof, although only exposed to part of the argument, if it 
 fits into our idea of the subject area, and if the details left to the reader seem standard, and so 
 ``robust'', in that the odd error should be easily fixable. Moreover an author's belief in their own work can be enhanced by examination and good  questions from colleagues: 
\begin{quote}
 There is no ... mathematician so expert in his science as to place entire confidence in his proof immediately on his discovery of it...Every time he runs over his proofs, his confidence increases; but still more by the approbation of his friends; and is rais'd to its utmost perfection by the universal assent and applauses of the learned world.    \\ \mbox{} \hfill ---  \textsc{   David Hume} (1739)
 \end{quote}
 Proofs are accepted as correct by  peers interested in similar questions and so  aware of the techniques used in the field.\footnote{Ideally a reviewer should be truly independent but in practice, for high-level research based on the literature, no one but an expert has the interest and skills to review the work, especially if it is complicated enough to require a substantial time commitment. Moreover an expert knows what is more-or-less standard in a new work, and what is novel, and so can better focus their efforts on where identifiable mistakes are more likely to appear.}
 To verify proofs they place the proof into a context they understand, their deep knowledge allowing them to skip and accept much tacit knowledge, while verifying the (relatively few) details that are different from what has gone before. The peer wishes to add to their own intuition and scope, not simply  agree that the proof's argument is correct.
The reader is not passive. She  wants to understand, to  synthesize and to use the ideas in her own research.
     
 Different people get different things out of a reading and therefore a new research article can inspire new ideas in hitherto unforeseen directions. Even the same person can, at different times,  get different things from  reading an article, our understandings do change over time, sometimes even how we approach the whole area. 
Proofs  accepted by these community standards  might be wrong since   the details are not carefully checked by the verifier (and indeed, most details are usually of less interest as an experienced reader can reconstruct them)  but the verifier is not looking for strict proof but rather understanding that allows the new work to be contextualized.\footnote{And slip-ups can occur: In \cite{Gr} we discuss some famous mistakes in world-leading mathematical research that were accepted by the community for a while, including works of Voevedsky, Biss, Mochizuki and, to a lesser extent, Wiles; and how to understand this acceptance in terms of what are the accepted standards of proof.} Proof and understanding are not synonymous, and we have to appreciate and  accept how they match and how they differ.   

A piece of mathematics feels right if  it is about what ``ought to be'' \cite{Chen}, rather than ``useful, fun, intriguing, beautiful, proved in detail'', a perspective which motivates the approach that many take to proving theorems. Indeed Paul Erd\"os claimed that an objective supreme being has a ``Book'' which contains the perfect proof for every true theorem, each of which is short and elegant. Short, so it is easy to verify, and elegant so one knows that the statement fits so well that it must be true.

Part of refereeing is to decide on the interest of the submission. Venkatesh  \cite{Venk} argues that ``the value we assign to a work of mathematics is purely subjective, in the sense that it depends solely on the perception of that work, and not on any objective quality''. Mathematics has long been directed by this kind of subjectivity and famously, Littlewood would ask referees, ``Is it new?   Is it correct?   Is it surprising?''
These questions are super-important since jobs, status and grants depend on where an author's work in published. Can there be an objective standard to judge new work, or must we maintain our subjective choices?
Can proof verification systems imitate (or improve on) human judgement?
   
\subsection*{The robust nature of proof}

We would like a review system in which the proof guarantees the theorem, so a competent mathematician does not need to verify each proof she uses herself.   Familiarity simplifies scrutiny (at some risk of favouring conformity) so papers are more quickly and accurately refereed,   focusing on those details that might be most likely, in the experience of the referee, to cause concern.  This supports a belief in the \emph{robust} nature of proof; we believe that  not much   can go wrong with well-used technical tools and so we make assumptions about what needs verifying.
And even if there is a mistake, experience shows that a simple modification should be enough to make the argument work. Any experienced researcher   does this regularly in developing their own work, so when they encounter minor technical flaws in the work of others, they tend to believe that they are fixable.


\subsection*{Commonalities between formal and intuitive proofs} 
All proofs are constructed in a similar way: Starting from agreed upon axioms we construct proofs of   given statements. To advance far we need to avoid going back to the axioms all the time, so we need to build a    library of statements that we know to be true and are unambiguously stated.
Traditionally this library is stored in research articles, and synthesized in books. 
Machines can also store this in their own language(s). 

A reader hopes to trace a thread back to the point she believes a claim with no further explanation. This can be achieved through
interactive links that allow a reader to dig progressively deeper, whether in a formal or intuitive proof.\footnote{Patrick Massot \cite{Mass} recently announced that software tools are   being developed to automatically convert formal proofs into such human-readable interactive proofs.} 

Naively this discussion suggests that a researcher can make advances through logical deductions from what they quote from the  library of known results. However researchers typically build awareness of their subject through speculation and proof, and seeing new questions in context, the  exact statement of already-known results playing a  subsidiary role.

\subsection*{Very different processes}

Formal proofs chase the details of  a proof back to the axioms, like a child tirelessly asking ``Why?'' (until one gets back to immutable truths). But, at the end of that process, does the child remember what they asked at the start and how they got to the end?   And when the formal proof verifier reports that a proof is correct then should we have any more belief than the child who is told, ``Because I told you so''?

Community proof verifiers typically learn from their work, and can not only reproduce the proof (in principle), but can  adapt the learned ideas to other questions. Participants in a community expect   proofs they can understand, interpret, appreciate, and even use if possible. They can  be excited to find an alternative or  clearer proof, though that seems to play no role in a formal system.


\subsection*{Eternal truths?} 

Venkatesh \cite{Venk} states that ``a proof is \dots an argument compelling consensus'' and de Toffoli \cite{DeT2}  that ``Shar[ing] mathematical arguments ... [is] a necessary condition of mathematical justification''. But why do we need to share if proofs build solidly from the axioms to the latest theorem?
De Toffoli \cite{DeT1} argues that 
 ``criteria of acceptability for rigorous proofs are not carved in stone ... but  are indexed to a mathematical community in a particular time'', citing proofs in basic calculus. This idea that proofs, and every question that we ask, belong to the latest paradigm, has long been recognized:
\begin{quote}
A triangle. This seems to be extremely simple, and you'd think we \dots know all about it\dots\,  Even if we prove that it possesses all the attributes we can conceive of, some other mathematician, perhaps 1000 years into the future, may detect further properties in it; so we'll never know for sure that we have grasped everything that there is to grasp about the triangle. 
\mbox{} \hfill ---  \textsc{  Ren\'e Descartes} (1648) 
\end{quote}
These thoughts make little sense if proofs and understanding are immutable, especially once the steps have been objectively verified so that they work in every context.   Thus formal proofs arguably are more useful/eternal than intuitive proofs only if mathematics deals in objective truth.

\section{Myths of scientific objectivity} 
Copernican theory was long objectively refuted by passages from the bible. Racism and sexism long supported and sustained by supposedly objective societal evidence.  Yet today those ``objective stances'' are, objectively, subjective and self-serving (see \cite{Hara} for much more on this theme). 
But surely in mathematics, objective proof follows from well-formulated axioms, language and deductive rules?\footnote{Indeed, in 1895, Peano observed that ``Imprecise ideas cannot be represented by symbols'', inferring that if we appropriately use symbols then everything will be precise and resulting truths will be objectively deduced. However even if we agree that this is a way to be precise, one can be precise and wrong!}
Calling an assertion objectively true gives it authority, it is rhetoric designed to quell doubt but how is it backed up?  Is there a non-subjective definition? (That is, uninformed by other human aspirations) If not, are such claims of certainty truly desirable?\footnote{``Best practice'' is a phrase currently used by bureaucrats to passively aggressively shame doubters. Is ``objectively true'' any better?}
What are the key components of objectivity?
\begin{quote}
Scientific objectivity \dots  expresses the idea that scientific claims, methods, results, and scientists themselves\dots should not be influenced by particular perspectives, value judgments, community bias or personal interests.\footnote{For this quote and a more thorough and less opinionated version of the ideas in this section, see the Stanford Encyclopaedia of Philosophy \cite{RS}.}
\end{quote}
How attainable is objectivity in any science?

Karl Popper defined ``objectivity'' as faithfulness to facts, interpreted without the bias of a premeditated perspective,  with the ability to adjust to changing information. However this does not mean that objectivity necessarily correctly identifies eternal truths.  He claimed that theories gradually come closer to truth, depending on our better collection and understandings of data, becoming more objective over time.  Thus, to Popper, objective truth is always something in front of us, to be strived for, but not objectively attainable.

Scientific realism claims that objective  truths are independent of perception,\footnote{A room has a certain temperature, whether or not you find it hot or I find it cold} though Hilbert's plurality of possible axiomatic bases (and alternative definitions given in mathematical works) suggest that many different perceptions are equally valid!  Even supposing we begin with the same axioms and definitions, each researcher's language use and viewpoint may lead to subtle differences with equally valid interpretations (which may be complementary, but it can be difficult to be certain). Anyway, who decides whose perception is objectively correct?

Kuhn \cite{Kuhn} was particularly skeptical of claims of objectivity, claiming that all observations are made through the lens of a reigning paradigm, 
perceived and conceptualised by the latest (but not last)  theoretical assumptions.\footnote{But surely experimental evidence is objective? However only   if the experimental apparatus is reliable, and typically  we only believe it's reliable if it confirms results obtained in line with our earlier beliefs.  Thus experimental evidence is only  understood from a perspective within its own paradigm.}
 The paradigm  guides  scientists' approach to their work, setting the community standards.\footnote{This can be consistent with Popper, for example in considering the popular example of special relativity enhancing and  supplanting Newtonian mechanics.}  This viewpoint  makes it hard to believe in a theory-independent language, observations or even objectivity; however   the concept of objectivity is arguably meaningful within a peculiar paradigm.
 
 One might argue that precision, simplicity, coherence and scope of an experiment or theory are value-free, and so \emph{should} motivate preferences.
However if we had written accuracy, elegance, fit to theory and connections to other parts of  the theory then 
these concepts are all inescapably embedded in some paradigm.

There have been massive paradigm shifts in mathematics in the last fifty years,   in the value placed on different areas,  stemming from the powerful influence of Groethendieck. His re-think of algberaic geometry, functional analysis and much else led the mathematical establishment to pay little attention to extraordinary work in any field not closely aligned with Groethendieck's work. For example combinatorics was trivialised as   ``mere calculation'' (as I heard Atiyah proclaim in the late '90s)\footnote{Indeed, shamefully,  Szemer\'edi was not  recognized in the general mathematics community for his great 1975 work on structure in non-sparse sets  \cite{Sze} until the 2012 Abel prize.} and it took a long time, and substantial controversy, before a Fields' medal was once again presented to research outside a narrow band of topics.

For what its worth, mathematical objectivity has suffered less from contextual influences than other sciences from pressures like climate denial, general politics, and needs of large businesses such as drug and food companies, though there are notable recent examples like  cryptography in industry and government, interpretation  of statistical data during the pandemic, and now the sometimes exaggerated claims of the machine learning industry.

\subsection*{Difficulties of definitions}
No single definition is useful for many of the concepts studied in modern mathematics, but all should be reconciled.\footnote{We are used to this in, for example, defining real numbers, or synonymous concepts in algebra and geometry.}
This can be an issue for formal verifiers especially as in cutting edge areas, many competing approaches (including definitions) can be tried, and only experience will determine which works best most often (and those judgement calls are inevitably subjective).

All formal languages must make choices as to how to explain ideas, meaning some thoughts are emphasized, others de-emphasized, even excluded.\footnote{This is also true of spoken languages. People who speak two languages find that some concepts are more naturally explained in one, some in the other.}  Therefore some new developments are easily expressed and worked on, others less so. That is a lot to reconcile with the notion of objectivity.

\subsection*{The uncertainty principle of objective proof verification} The history of mathematical practice suggests that \emph{the less one questions a proof,  the more susceptible it is to error.}
This  important principle strongly suggests  one must find a wide variety of ways to explain and to verify any given proof, even a computer proof, and to look at it from as many different perspectives as possible.

\section{Computers and proofs}

 There are currently three main uses of computers in proofs:\smallskip
 
$\bullet$ \emph{Calculations in establishing a proof}:  Authors might reduce their question to a large finite  problem with too many cases to resolve by hand, and then eliminate the cases by computer (like the proofs of the four color theorem (4CT), Kepler's conjecture for 3-d sphere packing, God's number for Rubik's cube, and the classification of finite simple groups). Authors might need to construct a tool with delicate special properties that might only exist in high dimension, and this can only feasibly be found by computer (like Maynard's 400-dimensional sieve in his work on short gaps between primes). Authors can use computers to calculate large examples which may inspire understanding and then proofs; this is the traditional role of papers in \emph{Mathematics \underline{of} Computation}, and we have seen some spectacular recent examples using machine learning (see, eg \cite{Dav}\footnote{Though beware of the hype, for example the title, ``guiding human intuition with AI'' which is a self-congratulatory way to describe, and purloin, the main role of computers in mathematics for the last 50 years.}).
  
$\bullet$ \emph{Assistance in verifying the logic of an author's arguments, perhaps interactively, ``computer-assisted proofs''}:  This has been most useful in verifying long, detailed proofs, where angels fear to tread.  For example, a team worked with Gonthier to verify the most recent proof of 4CT (which has 32 steps involving 633
subgraphs that need to be reduced) using the Coq v7.3.1 proof assistant;  a team worked with  Hales to verify his   proof of Kepler's conjecture\footnote{The original proof was so complicated that a team of a dozen referees worked for four years on verifying the proof and only reported they were ``99\% confident'' that the proof is correct.} using the HOL Light and Isabelle proof assistants.  
 
There have been striking recent advances with Lean, most notable  the verification, correction and improvement of a key result of Clausen and Scholze, discussed elsewhere in this volume. 
There are now a lot of people working with Lean, and the quality and quantity of ideas that have been codified is extraordinary and many areas of mathematics are seeing non-trivial results being verified. Moreover, as more researchers contribute to the system, interaction should move towards something resembling the high-level practice of mathematicians. With further work on the input and output languages, a system like this could be  user-friendly and  become an integral part of the mathematician's arsenal.

These proof-verifiers are  interactive, the user  interactively breaking the proof down into simpler objects that the machine already knows about, and  giving Lean hints.   
The proof assistant will determine whether the statement is `obviously' true or false based on its current library. If not, the user enters more details. The proof assistant therefore forces the user to explain their arguments in a rigorous way, and  to fill in simpler steps than human mathematicians might feel they need.\footnote{Scholze and Gonthier both reported that they   learnt a lot during this laborious input process!}
For example for a proof with ten lemmas, some the theorem-prover will see and resolve quickly, others it might need more details until it can see its way to a proof. In so-doing the program learns more,  maintains a library and is perhaps more efficient when it next encounters similar issues.

Proof assistants can't judge whether a mathematical statement is  interesting or important,  only whether it is consistent with what it has been shown.  They should eventually (maybe even soon) require less help, perhaps much less help.   The eventual proofs are not human readable, but thanks to
 Massot and his team \cite{Mass} that should change soon, converting machine proofs into high-level arguments that humans understand and appreciate, which allows us to have more faith in the output. Nonetheless, can we really trust proof-assistants to be correct  if they are only really checked by their own internal logic (which might propagate a subtle error)?  What would be best is if these proofs can be independently verified, perhaps by different programs run on different machines;  in effect, we propose refereeing computer proof verifiers output within their own community!\footnote{This will require them to share a common language.}   We can design the future based on what already works.

\medskip
 
$\bullet$ \emph{Proving claimed theorems,   ``computer-generated proofs''}:
  ``Machine learning'' typically develops its understanding in  simple ways as a result of clever algorithms.
Creating a large database and analyzing it with specially formulated tools can be startlingly effective (like Google Translate or ChatGPT) but this is not the same as developing intuition (or even simulating intuition effectively).\footnote{There is a lot of money and publicity surrounding the subject of ``machine learning'' and some other forms of ``artificial intelligence'' but   that many hyped advances are either  exaggerated or easily explained in terms of well-designed algorithms and extraordinary computing power.}  There are as yet no ``thinking machines''.

Relatively little is understood about creativity and intuition, and how humans move from one understanding to a rather different one. To simulate this on a machine seems very far away.\footnote{See Melanie Mitchell's wonderful book \cite{Mit} for a forensic discussion of what underlies some of recent developments in machine learning.}
 Before computers, librarians were often credited with knowing a lot more than they really did (as the gatekeepers of so much knowledge). Computers are  much bigger repositories for knowledge, more accessible and less proscribed by others, and can achieve some surprising feats; it is not surprising that they get credited with powers that they do not yet possess.

Computers have a large memory, whereas humans have to avoid severe combinatorial explosion by bringing in ``tactics'' early on in pruning search trees. For examples, a human might
\smallskip

--- Recognise which hypothesis in a statement is likely to be  most important;

--- Identify a subsidiary goal to aim for to break the proof   into smaller steps;

--- Identify and adapt a known proof technique  to this  new situation;

--- Use a few examples to guess at properties,  rapidly pruning  the search tree.
\smallskip

\noindent All of these are difficult to implement on a computer, so what tactics best serve computer generated proofs?
  Ganesalingam and   Gowers \cite{GG}   aimed to design a computer verifier to learn and think like a human,
 asking which order   to try  different tactics in, whether   computers can learn from their past experience, and whether we can devise a theory that better mimics human choices, or work with a mix of the two.\footnote{And see Gowers' latest project at https://gowers.wordpress.com/2022/04/28/announcing-an-automatic-theorem-proving-project/$\#$more-6531 .}
 One goal is to adapt and win Turing's ``imitation game'' \cite{Tur1}, but now a machine created proof should be indistinguishable from a  great human proof.\footnote{In  \cite{GG}, Ganesalingam and   Gowers selected problems to prove and got thousands of independent readers to try to distinguish which proofs were by their program and which by real people  (see https://gowers.wordpress.com/2013/04/14).
 The results are encouraging, though of course this is not the Turing test since it is not an independent arbiter that selected the problems.}

\subsection*{Computer errors} If formal proofs on a computer are to be objectively true then computers should become error-free.
People often assume that if a computer program   is reliable then it is ``free from error'' which flies in the face of their own experiences.  Reliable systems, which have been heavily invested in, govern your credit card, phone, airlines, your university, and we have all had frustration with those! Besides hardware and software problems, there may be programming issues (eg not accounting for your particular situation) or implementation issues (eg ambiguous menus, or misleading instructions). Surely over time these can all be ironed out, and computers will be trouble free?
 However upgrades might deal with a previously unmanaged case, but also provide new functions, which   bring new issues!
 And if a system eventually becomes ``perfect'', how would we prove it?

Computer hardware glitches are  usually recognized by  outputs that are obviously inconsistent with other information. However this does not help us when the computer is calculating something that we know little about!  Manufacturers  rarely reveal concerns about their products (so as not to put off potential purchasers), so if there is a glitch, it is unlikely to be  shared with users.
Moreover  would a small mathematical  error in a widely used computer chip be worth the cost of fixing for the manufacturers?\footnote{In 1993 Pentium released a chip which they subsequently found had a hardware bug   affecting its floating point processor (even when dividing certain seven digit integers by each other). Rather than recall the flawed chips they kept quiet and corrected the problem in updates. However Thomas Nicely made the error public in June 1994 after he found it when he got  a number theory calculation wrong.    Pentium resisted a recall until IBM refused to ship their product.}

Commercial software has about 1 bug per hundred lines of code; even the space program can only get it down to perhaps 1 per 10,000 lines. Corporations  keep quiet about bugs  to limit legal liability, and  correcting bugs can  creates new bugs!  There are recent program-verifiers designed to check that programs perform their claimed calculations so there may soon be less errors, but that remains to be seen.

\section{All the proofs yet to come...} 

Other articles in this issue address what theorems are to be predicted by computers, and even general intellectual leadership from computers, but such speculation is beyond my scope. Right now computer proof-verification and generation are fast-moving subjects and it is exciting to witness the beginning of these important developments.
Brilliant people are getting involved and Kevin Buzzard takes the compelling view that, ``the  more people are familiar with the software, the sooner interesting things will happen'' ---  Lean now has thousands of users.

My concern has been with the nature of proof and what it will become, largely focusing on refuting the naive notion that formal proofs will improve objectivity. But more important is that we should not lose the benefits of the community based approach to proof that has long served us so well. Indeed some of our traditions should guide our quest in developing computer proofs.

Most importantly, computers should serve our needs. Input should be in the standard lexicon, proofs should be presented to be human-readable, and verifiers developed to work alongside a research mathematician, giving her the ability to much more easily check out a proposed proof idea.  When a program claims to have a proof then it should be verified by 
different computers run by different chips, using different software.\footnote{Then it would be highly unlikely to get an error at the same point in the ``proof'' on each system, though this is still no guarantee of correctness.}    

Formal proofs are remarkably fragile (in that if we find any errors then it puts much more into doubt), and so we need to build in the robustness of traditional
intuitive mathematical proofs so that minor flaws can be fixed,

 Eventually we would like informative proofs, not just verifications,  \emph{explaining} the mathematics,  not necessarily give the shortest  proof.
 This should allow us to modify what we learn and help us to develop new concepts. We should be able to interact with the program as one does with a (marvellously retentive) human. It can be desirable to give several proofs for the same theorem, for example to highlight different themes -- can a computer program be trained to identify such variants?\footnote{One can ask whether   a  formalization of a given intuitive proof is going to be the same proof? When an intuitive proof is dissected into what is required for, say, Lean to work with, it will look very different and rest on a rather different looking library of knowledge.  And how different will the same proof look when modified for a different language?}
Is hard to describe what a good proof is (and thus Erd\"os's ``Book'') but we know it when we see it: Can  our computer programs   always find such proofs? 

I am asking for a lot, but now is surely the time to identify how to best direct these developments to enhance the future of proofs in mathematical research.

 \bibliographystyle{plain}

\end{document}